\documentclass[%
 reprint,
 amsmath,amssymb,
 aps,prl,
]{revtex4-2}

\usepackage{graphicx}
\usepackage{dcolumn}
\usepackage{bm}

\begin{document}

\preprint{DHpaper}

\title{Dirac-Harper Theory for One Dimensional Moir\'e Superlattices}

\author{Abigail Timmel}
\author{E. J. Mele}
\email{mele@physics.upenn.edu}
\affiliation{
 Department of Physics and Astronomy\\
 University of Pennsylvania \\
 Philadelphia, PA 19104
}

\date{\today}

\begin{abstract}
We study a Dirac Harper model for moir\'e bilayer superlattices where layer antisymmetric strain periodically modulates the interlayer coupling between two honeycomb lattices in one spatial dimension. Discrete and continuum formulations of this model are analyzed. For sufficiently long moir\'e period the we find low energy spectra that host a manifold of weakly dispersive bands arising from a hierarchy of momentum and position dependent mass inversions. We analyze their charge distributions, mode count  and valley-coherence using exact symmetries of the lattice model and approximate symmetries of a four-flavor version of the Jackiw-Rebbi one dimensional solution.
\end{abstract}

\maketitle

Lattices with competing periodicities are interesting as discrete systems poised in between primitive crystals and random networks \cite{Simon82,Ostlund84}.  When the ratio of the competing periods is a rational fraction $p/q$, the system is a crystal, although with superlattice translations that are inflated by the divisor $q$.   For these systems a spectral analysis of propagating Bloch waves is always possible, but it is ineffective if $q$ is large.  Famously, the related problem in {\it two dimensions}  has risen to prominence in addressing the behavior of moir\'e superlattices in bilayer graphene and twisted stacks of other two dimensional materials  \cite{Bistritzer,KangVafek,KoshinoTBLG,Po,Song}. In these situations, competing periodicities are introduced by a slight rotational misalignment between neighboring layers. For twisted bilayer graphene, important physics occurs near charge neutrality for rotation angles $\theta \sim 1^{\rm \circ}$ where as a practical necessity, one frequently adopts a continuum description \cite{Bistritzer}. This is a first-order-in-derivatives expansion about the microscopic Brillouin zone corners, where the microscopic lattice structure appears only through a spatial modulation of matrix-valued coupling between the Dirac operators of individual layers.  The correct interpretation of the narrow bands  at small twist angles in these models, as well as the analytic structure of its moir\'e-Bloch bands at general angles, remains the subject of a vigorous discussion \cite{KangVafek,KoshinoTBLG,Po,Song}.

In this Letter we analyze the one dimensional  lattice variant of the twisted bilayer problem using graphene as the prototype. The models we construct are one-dimensional in the sense that they have layer-antisymmetric strains which produce moir\'e-scale modulation in only one spatial dimension, retaining lattice-scale periodicity in the orthogonal coordinate. With a conserved crystal momentum, the problem  takes a strictly one dimensional form whose superlattice period is tuned by the lattice strain. This model generalizes the Harper equation \cite{Harper} for a scalar field $\psi_m$ in a one dimensional potential with period $1/\alpha$
\begin{eqnarray}
t \psi_{m+1} + t \psi_{m-1} + 2 \nu \cos \left( 2 \pi \alpha m  - \phi_o \right)  \psi_m = \epsilon \psi_m
\end{eqnarray}
by promoting $\psi_m$ to a four-component spinor ${\boldsymbol{\psi}}_m$ and the coupling parameters to matrices ${\boldsymbol{t}}(k_y)$ and ${\boldsymbol{\nu}}(k_y)$. We find that this modified Dirac-Harper (DH) model hosts narrow band physics near charge neutrality for sufficiently large moir\'e periods {\it without} fine tuning to a magic angle condition as is required in two dimensions. The near-zero mode structure originates from a reconstruction of the low energy spectrum through a {\it hierarchy} of mass inversions necessitated by the matrix structure of the interlayer potential, reminiscent of the multi-flavor  variant of the Jackiw Rebbi model\cite{Wen}.  Importantly, unlike its continuum counterpart, the DH model is formulated on a lattice and properly treats the coexistence of the discrete lattice and the moir\'e periods. The DH model does not admit a sharp identification of a conserved valley degree of freedom. Instead spectral degeneracies and topological properties of the bands are identified through the action of unitary and antinunitary symmetries on the discrete lattice Hamiltonian. Indeed we find that valley degeneracy, presumed to exist in continuum models, is generically preempted by an ordering field that is required by the lattice symmetry and is essential for understanding responses that are activated by the chirality of a bilayer.

\begin{figure}[!h]
	\begin{center}
		\includegraphics[width=\columnwidth]{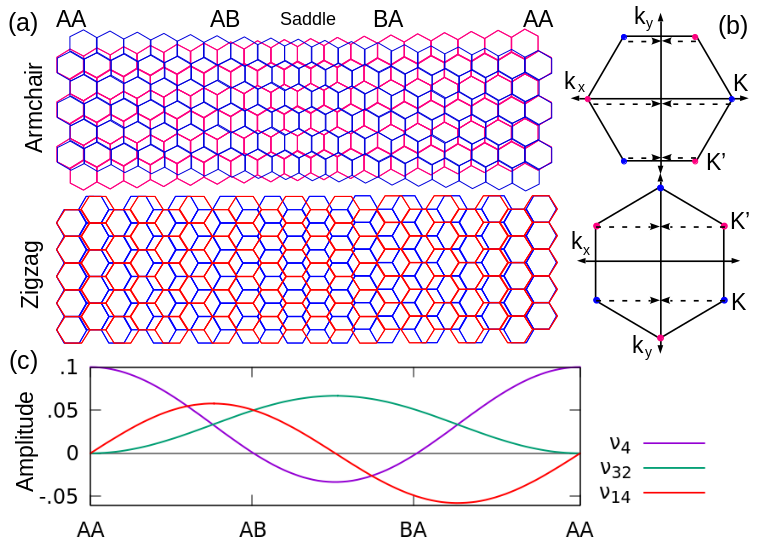}
	\end{center}
	\caption{(a) Moire graphene bilayer, varying smoothly between $AA$, $AB$, $BA$ and $SP$  stacking via layer antisymmetric shear (top) or uniaxial (bottom) strain.  (b) The orientation of the Brillouin zone controls the projection of $K$ and $K'$ points under zone folding, maximally separated for ZZ and mixed for AC.  (c) Amplitudes of the coefficients associated with each mass matrix across the moir\'e.}
	\label{structures}
\end{figure}

One-dimensional bilayer moir\'e patterns evolving from fully eclipsed ($AA$) through both staggered ($AB$, $BA$) stacking configurations are produced by layer antisymmetric strains (Figure~\ref{structures}a).  Two limiting cases are the armchair setting (AC, top) with shear strain and the zigzag setting (ZZ, bottom) with uniaxial strain.  The lattice orientations control the projection of the microscopic $K$ and $K'$ points onto the long axis of the folded moir\'e Brillouin zone, aligning them for AC and keeping them maximally separated in ZZ (Figure~\ref{structures}b).  Most of our analysis uses periodic boundary conditions on a large single moir\'e supercell, restricting $k_x=0$.

The Harper model for bilayer graphene generalizes Eqn.1
\begin{equation}
	\label{harper}
	{\bf t}(k) \boldsymbol{\psi}_{n+1} + {\bf t}^\dagger (k) \boldsymbol{\psi}_{n-1} + ({\bf t}_0 + \boldsymbol{\nu}(x))\boldsymbol{\psi}_n = \epsilon \boldsymbol{\psi}_n \\
\end{equation}
where ${\boldsymbol{\psi}}$ are four component fields with the conventional ordering of amplitudes on two sublattices and two layers:  ${\boldsymbol{\psi}}^T = (a_1,b_1,a_2,b_2)$.  Here ${\bf t}(k)$ are block diagonal $4 \times 4$  matrices derived from the nearest neighbor tight-binding kinetic energy for the individual graphene layers \cite{supplement} and $\boldsymbol{\nu}(x)$ is a spatially modulated interlayer coupling.  Defining the Pauli matrices $\sigma_i$ and $\tau_i$ acting on sublattice index and layer index respectively, the products $\sigma_x\tau_0 = \gamma_{15}$ and $\sigma_y\tau_0 = \gamma_{25}$ form a basis for ${\bf t}$
\begin{gather*}
	{\bf t}_0 = t\gamma_{15} \quad {\bf t}(k) = t_{15}(k)\gamma_{15} + t_{25}(k)\gamma_{25}
\end{gather*}
where $t$ is the hopping amplitude.  Explicit formulae for the coefficients $t_{15}$ and $t_{25}$ are tabulated in the supplementary section, \cite{supplement}.

The interlayer coupling $\boldsymbol{\nu}(x)$ is represented as a Fourier interpolation between $AA$, $AB$, and $BA$ stacking \cite{supplement}.  The high-symmetry configurations admit a represention in terms of Dirac matrices $\gamma_4 = \tau_x\sigma_0$, $\gamma_{32} = \tau_x\sigma_x$, and $\gamma_{14} = \tau_y\sigma_y$
\begin{gather*}
	\boldsymbol{\nu}(AA) = t'\gamma_4 \quad \boldsymbol{\nu}(AB/BA) = t'/2 (\gamma_{32} \pm \gamma_{14}) 
\end{gather*}
where $t'$ is the interlayer hopping strength.  The matrix-valued Fourier coefficients are defined on this basis of mass matrices, allowing the interlayer coupling to be written
$
	\boldsymbol{\nu}(x) = \nu_4\gamma_4 + \nu_{32}\gamma_{32} + \nu_{14}\gamma_{14}
$ with amplitudes
\begin{gather*}
	\nu_4 = t'/3(1 + 2\cos{2\pi x/L}) \quad \nu_{32} = t'/3(1-\cos{2\pi x/L})\\
	\nu_{14} = (t'/\sqrt{3})\sin{2\pi x/L}
\end{gather*}
plotted in Figure~\ref{structures}c.  The $\nu_4\gamma_4$ and $\nu_{32}\gamma_{32}$ terms ($\nu_{14}\gamma_{14}$ term) are symmetric (antisymmetric) under $x \rightarrow -x$, sublattice exchange, and layer exchange.  The $x$ or $y$-directed choice of sublattice orientation assigns symmetries of $\boldsymbol{\nu}(x)$ to the mirrors $\mathcal{M}_x$, $\mathcal{M}_y$ for ZZ and the two-fold rotations $C_{2x}$, $C_{2y}$, $C_{2z}$ for AC.

Linearizing this problem in the low energy regime gives the continuum variant of the DH model:
\begin{equation}
\label{linearDH}
	(i \tilde{\bf t}_x \frac{\partial }{\partial x} + \tilde{\bf t}_y q_y + \boldsymbol{\nu}(x))\boldsymbol{\psi} = \epsilon \boldsymbol{\psi}
\end{equation}
where $\tilde{\bf t}_i$ expresses the expansion of ${\bf t} = {\bf t}_0 + {\bf t}(k)$ in small $q_i$ about either Dirac point where ${\bf t} = 0$.  This expansion is parameterized by a Fermi velocity $\hbar v_F = \sqrt{3}ta/2$  where $a$ is the microscopic lattice constant.  The $(\tilde{\bf t}_{x},\tilde{\bf t}_y)$ reduce to $\hbar v_F(\gamma_{15},\gamma_{25})$ for AC and $\hbar v_F(\gamma_{25},\gamma_{15})$ for ZZ \cite{supplement}.  The continuum model requires a choice of valley ($K$ or $K'$) about which to carry out the expansion, and consequently it implies a spectral doubling associated with the conserved valley index.  Although this simplification enforces a valley polarization which is not present in the discrete Dirac-Harper model, it can be useful for analytic purposes.

\begin{figure}[!h]
	\begin{center}
		\includegraphics[width=\columnwidth]{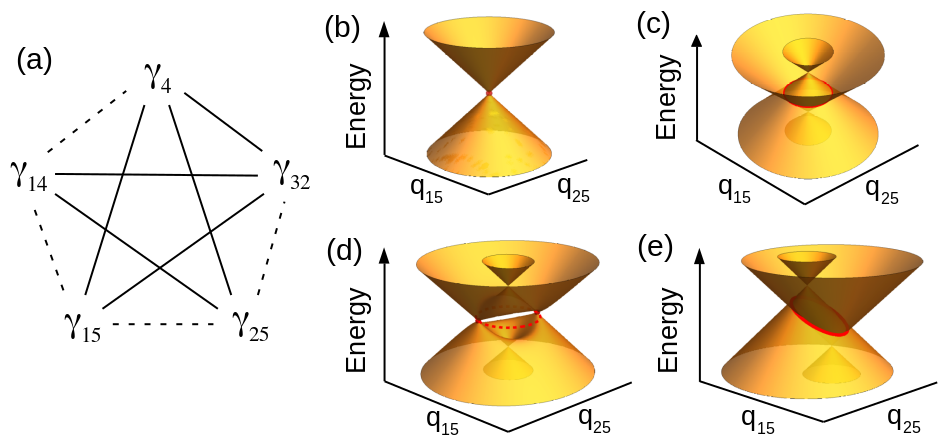}
	\end{center}
	\caption{(a) $K_5$ graph representing commutation relations of the five Dirac matrices appearing in the Hamiltonian.  Commuting pairs are connected by a solid edge and anticommuting pairs by a dashed edge. (b) Dirac point in the absence of any mass terms, where $q_{15}$ and $q_{25}$ refer to the momenta paired with $\gamma_{15}$ and $\gamma_{25}$ for a given lattice orientation.  (c) Cone splitting induced by $\nu_4$, widening the zero energy locus to a critical ring (in red).  (d) Gap opened by $\nu_{14}$ on the critical ring with point closures enforced along $q_{25}$.  (e) Displacement of cones along $q_{15}$ by $\nu_{32}$, tilting the critical ring.}
	\label{commutation}
\end{figure}

\begin{figure}[!h]
	\begin{center}
		\includegraphics[width=\columnwidth]{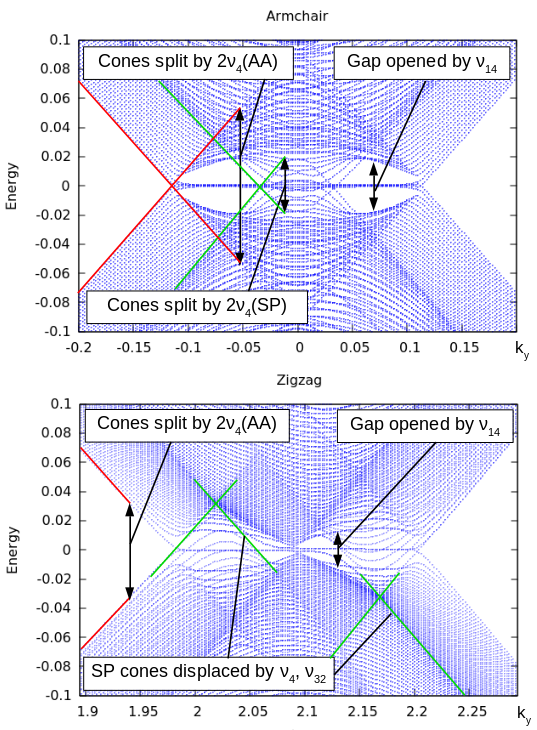}
	\end{center}
	\caption{Armchair and zigzag band structures calculated from the discrete DH model for moir\'e length 1800 times the lattice constant and $t/t' = 0.1$.  Spectra were obtained via direct diagonalization using GNU Octave \cite{Octave}.}
	\label{spectra}
\end{figure}

The electronic spectrum of a one-dimensional moir\'e superlattice in bilayer graphene has been studied previously  within a valley-projected continuum theory \cite{sanJose1} and in a lattice model \cite{Gonzalez} which focused on the non-Abelian character of the interlayer coupling matrices.  For sufficiently large moir\'e periods both models predict a manifold of narrow bands near charge neutrality but interestingly they have quite different underlying mode structures \cite{Gonzalez}.  This can be appreciated by comparing the spectra calculated from the discrete DH model (Equation~\ref{harper}) shown in Figure~\ref{spectra}. The differences arise from the commutation relations among the mass terms $\boldsymbol{\nu}(x)$ and the kinetic energy operators, represented in Figure~\ref{commutation}a as a $K_5$ graph with two edge types.  The AC and ZZ structures are distinguished by the relative roles of kinetic energy operators $\gamma_{15}$ and $\gamma_{25}$ carrying different commutation relations with the mass terms. These distinctions encode the different symmetries of these structures and persist even in the limit of large moir\'e period where they control the low energy narrow band physics.

The sublattice-diagonal $\gamma_4$ mass term commutes with both kinetic terms and displaces the layer-degenerate Dirac cones (Figure~\ref{commutation}b) by energy $\pm\nu_4$ (Figure~\ref{commutation}c).  These displaced cones intersect on a ring at $E=0$ of radius $\nu_4/\hbar v_F$ (henceforth called the critical ring) which defines the extent of a band-inverted region.  In the full moir\'e these widths are defined by the $\nu_4$ extrema in both the $AA$ and $SP$ regions, and the spectra in Figure~\ref{spectra} display projections of cones displaced on both scales.  The sublattice antisymmetric $\gamma_{14}$ mass term carries a different commutation relation with each kinetic term, distinguishing the two lattice orientations.  Anticommutation with both $\gamma_4$ and $\gamma_{25}$ opens a gap around zero energy, but commutation with $\gamma_{15}$ enforces a pair of gap closures on the $q_{15}$ axis (Figure~\ref{commutation}d) \cite{supplement}.  These closures both project to $q_y=0$ in ZZ of the full moir\'e, whereas for AC they are maximally separated on either side of the gap (Figure~\ref{spectra}).  The third mass matrix $\gamma_{32}$ plays a secondary role: it commutes with every term except $\gamma_{25}$, displacing the two cones in opposite directions along the $q_{15}$ axis (Figure~\ref{commutation}.  This is reflected in the $q_y$ translation of $\nu_4(SP)$ displaced cones in the ZZ spectrum (Figure~\ref{spectra}).

\begin{figure}[!h]
	\begin{center}
		\includegraphics[width=\columnwidth]{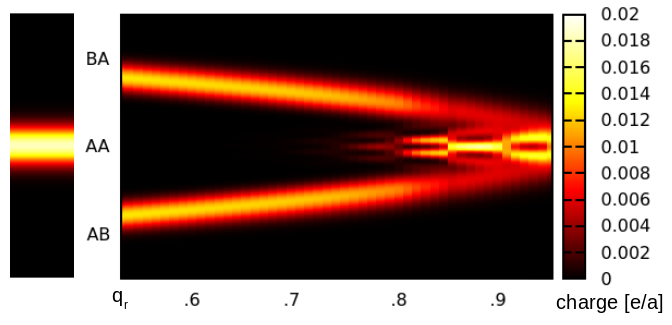}
	\end{center}
	\caption{Charge densities for $AA$-peaked modes (left) and $AB/BA$-peaked modes as a function of $q_r = k_y/q_c$ (right) where $q_c$ is the radius of the critical ring $t'/\hbar v_F$.  These are calculated numerically using the discrete DH equation.}
	\label{modes}
\end{figure}

In the gaps opened by $\nu_{14}$  we observe a manifold of weakly dispersive bands near zero energy.  Although the $\nu_4(SP)$ cones overlap this gap significantly in ZZ, they can be shifted in energy by tuning $\nu_{32}$.  For ZZ, pairs of nearly degenerate modes remain weakly dispersive over half the critical ring width and disperse with opposite velocities at the ring edges.  Two sets of these together span the full critical ring, and summing over the two inequivalent valleys we find a total of eight modes with charge densities confined to the $AA$ region (Figure~\ref{modes}).  In the case the AC moir\'e, pairs of nearly degenerate modes span the entire critical ring diameter and are exactly pairwise degenerate.  Two of these pairs (four branches) have charge densities peaked in the $AA$ region, and another two (four branches) have densities that shift from the $AB$ and $BA$ regions toward the $AA$ region as a function of $k_y$ (Figure~\ref{modes}).  The total mode count is again eight branches, though with twice the $k_y$ measure of the ZZ modes.

Exact symmetries of the two-valley discrete DH model can be utilized to study valley coherence.  Augmenting the $\boldsymbol{\nu}(x)$ symmetries to preserve momentum, these are $\mathcal{M}_x$, trivially $\mathcal{M}_y \mathcal{T}$, and their product $C_{2z}\mathcal{T}$ for ZZ and $C_{2x} \mathcal{T}$, $C_{2y}$, and their product $C_{2z}\mathcal{T}$ for AC \cite{supplement}.  The $C_{2y}$ symmetry of AC is notable because it exchanges valleys and thereby imposes a valley-mixed eigenbasis.  It does not commute with the coexisting antiunitary symmetries, enforcing twofold degeneracies.  From symmetry under valley exchange in $C_{2y}$, we can understand the multiplicity to be a consequence of valley coherence.  ZZ does not support this type of degeneracy with just one unique nontrivial symmetry operation, but we understand the analogous doubling of the energy eigenstates states to arise from its two momentum-separated valleys.

Low-energy approximate symmetries of the continuum DH model describe single-valley contributions to the mode count.  Defining a single-valley `time-reversal' $\mathcal{T}_K$ to reverse momenta $q$ about a Dirac point, the operator $\mathcal{P}_K = \tau_z\mathcal{T}_K$ negates the spectrum in a single-valley `particle-hole' symmetry.  In AC, $\mathcal{T}_K$ leaves the spectrum invariant due to the purely-imaginary kinetic terms, so the spectrum also exhibits chiral symmetry.  In ZZ, the purely real kinetic energy terms produce a modified spectrum under $\mathcal{T}_K$, breaking chiral symmetry.  Since the weakly-dispersing bands are not exactly at zero energy, this chiral symmetry doubles the mode count in AC, and particle-hole symmetry is consistent with the two sets of half ring-width bands found in ZZ.  More precisely, the spectra at $\pm q_{25}$ are related by the operator $C_{2x}$, which by commuting with all terms but $\gamma_{25}$, preserves the energies while reversing $q_{25}$.  In AC this manifests in the low bands extending across the full $q_y$ width of the ring, and in ZZ it causes a low-energy approximate degeneracy of states at momenta $\pm q_x$ on the critical ring.  Conversely, the sublattice-even $\gamma_{15}$ is reversed via $\mathcal{P}_K$ which also inverts the spectrum.  This causes pairs of states at $\pm q_x$ in AC and $\pm q_y$ in ZZ to lie at opposite energies, completing a measure of two full-width bands per valley.

The symmetries of the continuum DH model can be associated with an Altland Zirnbauer symmetry classification.  From the local Hamiltonian $\mathcal{H}(x) = \tilde{\bf t}_y q_y + \boldsymbol{\nu}(x)$, one can classify band manifolds by topological invariants computed at each value of $k_y$.  The ZZ Hamiltonian is a real operator which commutes with the complex conjugation operator $\mathcal{K}$ and therefore lies in class AI.   AC commutes with $C_{2x}\mathcal{K}$ and anticommutes with $\tau_z\mathcal{K}$, putting it into to class BDI.  At fixed momentum $k_y$, these one dimensional Hamiltonians depend on a single position-like coordinate $(x)$, so the bands of both structures are classified by a $\mathbb{Z}_2$ topological index \cite{Chiu}. In AC, the even multiplicity of states in isolated bands prevents juxtaposition of topologically distinct regions.  In ZZ, a loop integral of the connection $\langle u | \partial_x u \rangle$ over the moir\'e period in $x$ yields values $\pi$ inside and $0$ outside the ring boundary $q_y = t'/\hbar v_F$ \cite{supplement}.  These two sectors are isolated by the low energy modes that become strongly dispersive at the critical ring.

This analysis identifies the relation between the $AA$ peaked states found in both settings, but there exists a second set of four $AB/BA$ peaked states in AC which have no analog in ZZ.  The presence of spatially localized modes near zero energy suggest a Jackiw-Rebbi (JR) mechanism for a relativistic model with mass inversion \cite{JackiwRebbi}.  Indeed the $AA$ regions correspond to sign  changes of $\nu_{14}$, and the $AB$ and $BA$ regions present sign changes of a coupled mass term \cite{supplement}.  The four-component DH problem cannot be reduced to an exact first order single-component theory unless certain conditions \cite{JackiwRebbi} are satisfied, but for slowly varying mass terms we can diagonalize the continuum DH equation at zero energy
\begin{equation}
	\label{JReq}
	\frac{\partial}{\partial x}\psi = (i\tilde{\bf t}_x)^{-1}(\tilde{\bf t}_y q_y + \boldsymbol{\nu}(x)) \psi
\end{equation}
by ignoring the order $1/L$ commutator of $\frac{\partial}{\partial x}$ with diagonalization of the RHS.  The real part of the eigenvalues of the RHS determine evanescent growing and decaying solutions, and critical points correspond to the nullspace of
\begin{equation*}
	M(x) = \tilde{\bf t}_x q_x + \tilde{\bf t}_y q_y + \boldsymbol{\nu}(x)
\end{equation*}
i.e. the local Hamiltonian \cite{supplement}.  In AC, this nullspace is nontrivial in the $AA$ region and a pair of $q_y$-dependent points between $AB/BA$ and $AA$ where the mass terms locally support gap closures.  In ZZ, it is nontrivial in $AA$ and a range of points around $SP$ where the $SP$-displaced cones cross zero energy.  These points all offer eigenvalue branches changing sign which allow solutions to peak at the observed locations \cite{supplement}.

The one-dimensional moir\'e structures studied in this Letter support narrow band physics for sufficiently large superlattice periods without requiring fine-tuning to a magic angle condition.  An extension of the discrete theory to two dimensions can describe the competition between the lattice and moir\'e periods with physical consequences that are inaccessible to the continuum theory but are important for any response of the system that is activated by chirality.  The drift of the AB/BA peaks as a function of $k_y$ is reminiscent of  anomalous transport in the Hall effect, although time-reversal symmetry here forbids a charge Hall conductance.  Nevertheless, the breaking of $\mathcal{M}_x$ in the AC structure allows anomalous transport for a neutral transverse dipole current consistent with time reversal symmetry, the active point symmetries and the low energy mode count.

This work was supported by the Department of Energy under grant DE-FG02-84ER45118.

\nocite{*}

\pagebreak
\widetext
\begin{center}
	\textbf{\large Supplementary Information: Dirac-Harper Theory for One Dimensional Moire Superlattices}
\end{center}
\setcounter{equation}{0}
\setcounter{figure}{0}
\setcounter{table}{0}
\setcounter{page}{1}
\makeatletter
\renewcommand{\theequation}{S\arabic{equation}}
\renewcommand{\thefigure}{S\arabic{figure}}
\renewcommand{\bibnumfmt}[1]{[S#1]}
\renewcommand{\citenumfont}[1]{S#1}

\section{Lattice kinetic energy operators}

\begin{figure}[!h]
	\begin{center}
		\includegraphics[width=0.5\columnwidth]{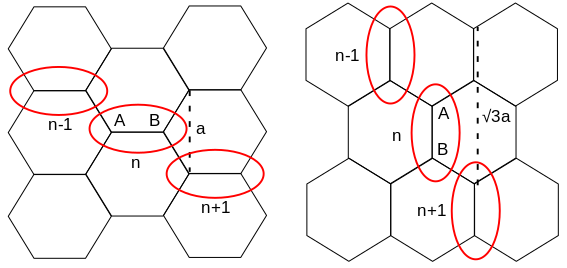}
	\end{center}
	\caption{Cell arrangement for ZZ (left) and AC (right).}
	\label{cells}
\end{figure}

The kinetic energy of the discrete DH model follows from the tight-binding model using a choice of cells shown in Figure~\ref{cells}:
\begin{gather*}
	T_{AC} = t\sum_{n,\tau} \psi_{A_n}^\dagger \psi_{B_n} + \psi_{A_{n+1}}^\dagger \psi_{B_n} + e^{-i\sqrt{3}k_y a}\psi_{A_{n-1}}^\dagger \psi_{B_n} + H.c. \\
	T_{ZZ} = t\sum_{n,\tau} \psi_{A_n}^\dagger \psi_{B_n} + \psi_{A_{n+1}}^\dagger \psi_{B_n} + e^{ik_y a}\psi_{A_{n+1}}^\dagger \psi_{B_n} + H.c.
\end{gather*}
where $\tau$ is the layer index, and $A$, $B$ are the two sublattice sites.  A Bloch phase of $e^{i\phi}$ where $\phi_{AC} = \frac{1}{2}(k_x L - \sqrt{3}k_y L)$ and $\phi_{ZZ} = \frac{1}{2}(-k_y L + \sqrt{3}k_x L)$ is acquired at the end of the moir\'e supercell.  It is useful to apply a gauge transformation $U = e^{i\phi \frac{n}{N}}\psi_n^\dagger\psi_n$ to distribute this phase accumulation evenly across the supercell, resulting in matrix representations
\begin{gather}
	\label{matrices}
	\text{AC}: \; {\bf t} = \tau_0 \otimes t\begin{pmatrix}
		0 & e^{(\frac{i k_x a}{2} - \frac{i\sqrt{3}k_y a}{2})}\\
		e^{(\frac{i k_x a}{2}+\frac{i\sqrt{3}k_y a}{2})} & 0 
	\end{pmatrix} \quad
	\text{ZZ}: \; {\bf t} = \tau_0 \otimes t\begin{pmatrix}
		0 & e^{\frac{i\sqrt{3}k_x}{2}}(e^{\frac{i k_y a}{2}} + e^{-\frac{i k_y a}{2}})\\
		0 & 0
	\end{pmatrix} \\
	{\bf t}_0 = \tau_0 \otimes t\begin{pmatrix}
		0 & 1 \\
		1 & 0
	\end{pmatrix} 
\end{gather}
for the Harper kinetic energy 
$${\bf t}\psi_{n+1} + {\bf t}^\dagger\psi_{n-1} + {\bf t}_0 \psi_n$$

These can be written in the basis $\gamma_{15}$, $\gamma_{25}$ with intracell term ${\bf t}_0 = \gamma_{15}$ and intercell ${\bf t}(k) = t_{15}\gamma_{15} + t_{25}\gamma_{25}$ whose coefficients are tabulated in Table~\ref{harper_coeff}.

\begin{table}[!h]
	\caption{Coefficients of $\gamma_{15}$ and $\gamma_{25}$ in the Harper equation for armchair and zigzag.}
	\label{harper_coeff}
	\begin{center}
		\begin{tabular}{l|ll}
			& $t_{15}(k)$ & $t_{25}(k)$ \\
			\hline
			AC & $t(e^{\frac{ik_x a}{2}}\cos{\frac{\sqrt{3}k_y a}{2}})$ & $t(e^{\frac{ik_x a}{2}}\sin{\frac{\sqrt{3}k_y a}{2}})$ \\
			ZZ & $t(2e^{i\frac{\sqrt{3}k_x a}{2}}\cos{\frac{k_y a}{2}})$ & $t(2ie^{i\frac{\sqrt{3}k_x a}{2}}\cos{\frac{k_y a}{2}})$ \\
		\end{tabular}
	\end{center}
\end{table}

This gauge has the advantage that there is no $k_y$ phase accumulation under operations taking $x \rightarrow -x$.  Such a phase accumulation occurs in the original gauge because a reversal $x \rightarrow -x$ shifts the $n^{\text{th}}$ marked cell by $y = \sqrt{3} n a$ in AC or $y = n a$ in ZZ, accumulating a phase $U' = e^{ik_y y}$.  The operator reversing $x$ then takes the form 
$$M_x = U'\psi_{n}^\dagger\psi_{N-n}$$
which transforms to the new gauge as 
$$U^\dagger M_x U = e^{-\frac{i k_y na}{2}} (e^{i k_y na} \psi_{n}^\dagger\psi_{N-n}) e^{\frac{i k_y (N-n) a}{2}} = e^{-i k_y N a} \psi_n^\dagger \psi_{N-n}$$
from which we can drop the constant phase to write $M_x = \psi_n^\dagger \psi_{N-n}$.

The symmetries of the kinetic term 
$$ {\bf t}(k)\psi_{n+1} + {\bf t}^\dagger\psi_{n-1} + {\bf t}_0\psi_n$$
can be written as combinations of $M_x$ (which effectively takes ${\bf t} \rightarrow {\bf t}^\dagger$), sublattice exchange $\sigma_x$, layer exchange $\tau_x$, and complex conjugation $\mathcal{K}$.  Since $\boldsymbol{\nu}(x)$ is invariant under any pair of $M_x$, $\sigma_x$, and $\tau_x$, we only consider operations with an even multiplicity of these operators. From Equation~\ref{matrices}, it is clear that $M_x \sigma_x \mathcal{K} \equiv C_{2z}\mathcal{T}$ which takes ${\bf t} \rightarrow ({\bf t}^\dagger)^\dagger$ is a symmetry for both structures.  Additionally, at $k_x = 0$, the AC kinetic terms are invariant under $\sigma_x \mathcal{K}$ and $M_x$, which can be augmented to symmetries of $\boldsymbol{\nu}(x)$ by combining with $\tau_x$ to yield $\sigma_x \tau_x \mathcal{K} \equiv C_{2x} \mathcal{T}$ and $M_x \tau_x \equiv C_{2y}$.  In ZZ at $k_x = 0$, $\mathcal{K}$ is trivial so we can split $C_{2z}\mathcal{T}$ into $M_x \sigma_x \equiv \mathcal{M}_x$ and the trivial $\mathcal{M}_y \mathcal{T}$.

\begin{figure}[!h]
	\begin{center}
		\includegraphics[width=0.4\columnwidth]{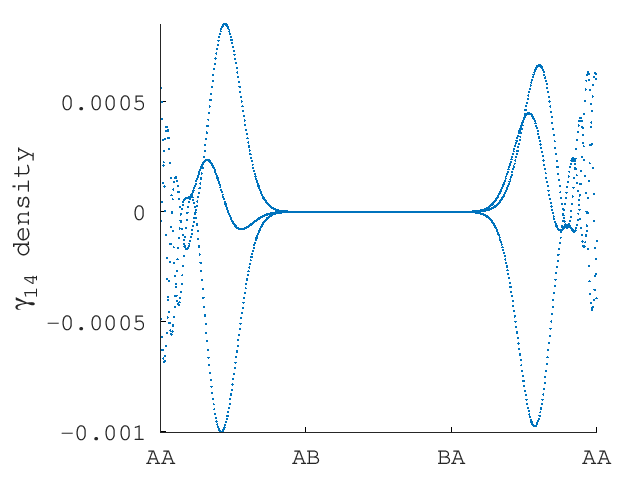}
	\end{center}
	\caption{Period-three oscillation observed in the $\gamma_{14}$ density at $q_r = .85$ where $q_r = k_y/(t'/\hbar v_F)$ is the ratio of $k_y$ to the radius of the critical ring.}
	\label{threefold}
\end{figure}

The gauge dependence of $M_x$ makes it easy to see why in AC $C_{2y}$ does not commute with the coexisting antiunitary symmetries.  If $M_x$ contains an $x$-dependent phase $U(x)$, commuting with $\mathcal{K}$ takes $U(x) \rightarrow U^*(x)$.  The presence of $C_{2y}$ guarantees that there exists a valley-mixed eigenbasis, which credits the multiplicity of the degeneracy to valley index.  Given that $C_{2z} \mathcal{T}$ is a simultaneous symmetry that preserves valley index, this does not exclude the possibility of a valley polarized basis.  Nevertheless, we find a period-three oscillation in the density of certain operators, $\gamma_{14}$ for instance (Figure~\ref{threefold}), provides evidence of valley coherence.  Therefore we can conclude that the valley degree of freedom is not an exact symmetry of AC.  At low energy it is a good symmetry of ZZ because the valleys project to different momenta $k_y$.

\section{Interlayer Coupling}

The interlayer potential is a Fourier series of the form
\begin{equation*}
	\boldsymbol{\nu}(x) = \boldsymbol{\nu}_0 + \boldsymbol{\nu}_{-1}e^{\frac{-2\pi ix}{L}} + \boldsymbol{\nu}_{1}e^{\frac{2\pi ix}{L}}
\end{equation*}
Using Fourier analysis to fix $AA$, $AB$, and $BA$ stacking at $x=0$, $2\pi/3L$, and $4\pi/3L$ respectively, the matrix-valued coefficients evaluate to 
\begin{gather*}
	\boldsymbol{\nu}_0  = \frac{t'}{3}(\gamma_{32} + \gamma_4) \quad \boldsymbol{\nu}_{\pm 1} = \frac{t'}{6}(2\gamma_4 - \gamma_{32} \mp i\sqrt{3}\gamma_{14})
\end{gather*}
The Fourier series with these coefficients can be rearranged into
\begin{equation*}
	\boldsymbol{\nu}(x) = \nu_4\gamma_4 + \nu_{32}\gamma_{32} + \nu_{14}\gamma_{14}
\end{equation*}
\begin{gather*}
	\nu_4 = t'/3(1 + 2\cos{2\pi x/L}) \quad \nu_{32} = t'/3(1-\cos{2\pi x/L})\\
	\nu_{14} = (t'/\sqrt{3})\sin{2\pi x/L}
\end{gather*}
which are reproduced in the main text.

\section{Continuum model}

To linearize the kinetic energy, we expand the kinetic energy ${\bf t} + {\bf t}^\dagger + {\bf t}_0$ in small momentum $q$ about a Dirac point, i.e. $(k_x,k_y) = (4\pi/3,0)$ for AC and $(0,4\pi/3)$ for ZZ.

\begin{gather*}
	\label{expansion}
	\text{AC}: \; \tau_0 \otimes t\begin{pmatrix}
		0 & 1 + e^{(\frac{i k_x a}{2} - \frac{i\sqrt{3}k_y a}{2})} + e^{(\frac{-i k_x a}{2} - \frac{i\sqrt{3}k_y a}{2})}\\
		1 + e^{(\frac{i k_x a}{2}+\frac{i\sqrt{3}k_y a}{2})} + e^{(\frac{-i k_x a}{2}+\frac{i\sqrt{3}k_y a}{2})} & 0 
	\end{pmatrix} \\
	\approx -\tau_0 \otimes ta\frac{\sqrt{3}}{2}\begin{pmatrix}
		0 & 1\\
		1 & 0
	\end{pmatrix} q_x - \tau_0 \otimes ta\frac{\sqrt{3}}{2}\begin{pmatrix}
		0 & -i \\
		i & 0
	\end{pmatrix}q_y = -\hbar v_F (\gamma_{15}q_x + \gamma_{25}q_y)\\
	\\
	\text{ZZ}: \; \tau_0 \otimes t\begin{pmatrix}
		0 & 1 + e^{(\frac{i k_y a}{2} + \frac{i\sqrt{3}k_x}{2})} + e^{(-\frac{i k_y a}{2} + \frac{i\sqrt{3}k_x}{2})}\\
		1 + e^{(\frac{i k_y a}{2} + \frac{-i\sqrt{3}k_x}{2})} + e^{(-\frac{i k_y a}{2} + \frac{-i\sqrt{3}k_x}{2})} & 0
	\end{pmatrix} \\
	\approx -\tau_0 \otimes ta\frac{\sqrt{3}}{2}\begin{pmatrix}
		0 & 1\\
		1 & 0
	\end{pmatrix} q_y + \tau_0 \otimes ta\frac{\sqrt{3}}{2}\begin{pmatrix}
		0 & -i \\
		i & 0
	\end{pmatrix}q_x
	= -\hbar v_F (\gamma_{15}q_y - \gamma_{25}q_x)\\
\end{gather*}

\section{Local spectra and the $\mathbb{Z}_2$ invariant}

From this linearized model, we can examine the energies under the addition of each term in $\boldsymbol{\nu}(x)$. Letting $q_{15}, q_{25}$ denote the components of $q$ paired with $\gamma_{15},\gamma_{25}$, the anticommuting kinetic terms yield the expected Dirac cone
$$ E = \pm v_F\sqrt{q_{15}^2 + q_{25}^2} $$
Adding a constant $\nu_4\gamma_4$ term, which commutes with both components of the kinetic energy, we encounter cones split by energy $\pm \nu_4$
$$ E = \nu_4 \pm v_F\sqrt{q_{15}^2 + q_{25}^2}$$
With a $\nu_{14}\gamma_{14}$ term, the spectra along the $q_{15}$ and $q_{25}$ axes become
\begin{gather*}
	E(q_{15}) = \pm \sqrt{(v_F q_{15} \pm \nu_4)^2 + \nu_{14}^2} \quad E(q_{25}) = v_F q_{25} \pm \sqrt{\nu_4^2 + \nu_{14}^2}
\end{gather*}
which in the first case describes hyperbolas opening a gap around zero energy, and in the second case is a cone split by $\sqrt{\nu_4^2 + \nu_{14}^2}$ enforcing point gap closures on the $q_{25}$ axis.  Finally, with $\nu_{32}\gamma_{32}$ 
\begin{gather*}
	E(q_{15}) = \nu_{32} \pm \sqrt{(v_F q_{15} \pm \nu_4)^2 + \nu_{14}^2} \quad E(q_{25}) = \sqrt{\nu_{32}^2 + (v_F q_{25})^2} \pm \sqrt{\nu_4^2 + \nu_{14}^2}
\end{gather*}

\begin{figure}[!h]
	\begin{center}
		\includegraphics[width=0.3\columnwidth]{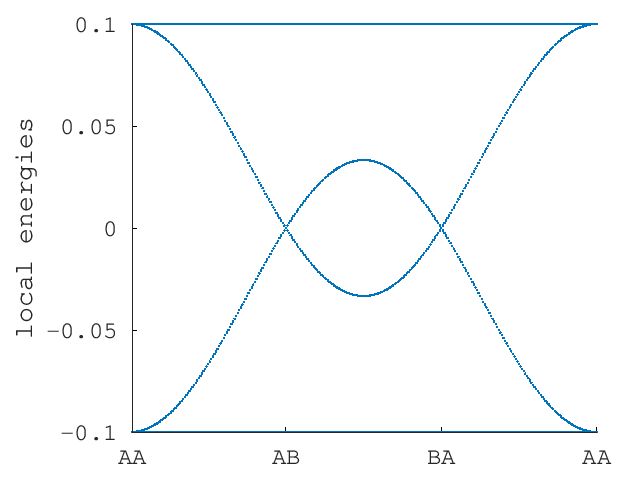}(a) 
		\includegraphics[width=0.3\columnwidth]{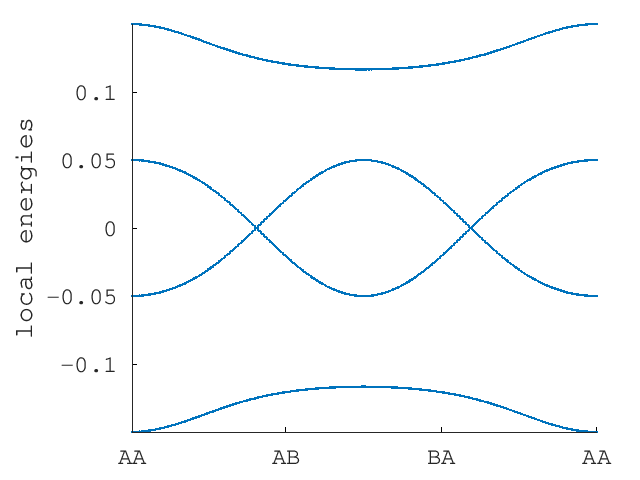}(b)
		\includegraphics[width=0.3\columnwidth]{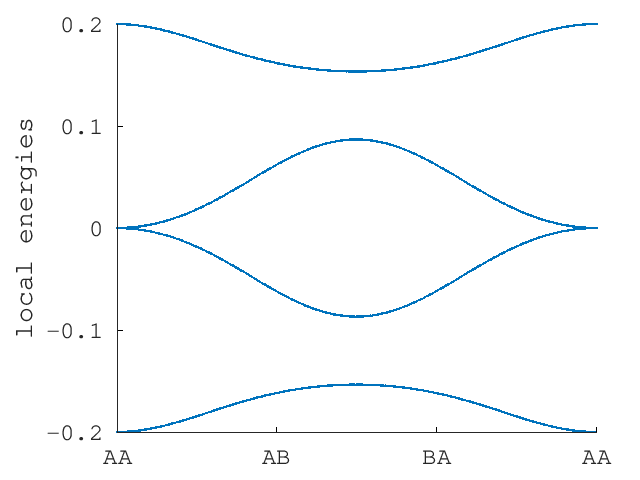}(c)
	\end{center}
	\caption{Local spectra at $q_{15}=0$ and (a) $q_{25} = 0$, (b) $q_{25} = .5 q_c$, and (c) $q_{25} =  q_c$ plotted as a function of moir\'e position.  This corresponds to AC at $k_x=0$ and shows the gap closure moving from $AB/BA$ stacking to $AA$ stacking as a function of $k_y$.}
	\label{AC_local}
\end{figure}

As $\boldsymbol{\nu}(x)$ modulates the amplitudes of the mass terms across the moir\'e, the magnitude of $\sqrt{\nu_4^2 + \nu_{14}^2}$ and $\nu_{32}$ changes as a function of $x$, moving the $q_{25}$ location of the gap closures.  Conversely, for every $q_{25}$ within the critical ring, there is some combination of mass amplitudes within the moir\'e where the spectrum admits zero eigenvalues.  These gap closures along the spatial dimension occur in the $AB/BA$ regions at $q_{25} = 0$ and move outward to the $AA$ region as $q_{25}$ approaches $q_c$, shown in Figure~\ref{AC_local}.  With explicit formulae inserted for the mass terms, this position is given by
$$ x = \frac{L}{2\pi}\cos^{-1}(\frac{3}{2}(\frac{v_F q_y}{t'})^2 - \frac{1}{2})$$
where $x=0$ is defined as the $AA$ region.  Another natural location for gap closures in the moir\'e is the pure $AA$ stacking.  Choosing $q$ so that $q_{15}^2 + q_{25}^2 = q_c^2$, the local mass configuration of the $AA$ region always supports zero eigenvalues.

\begin{figure}[!h]
	\begin{center}
		\includegraphics[width=0.3\columnwidth]{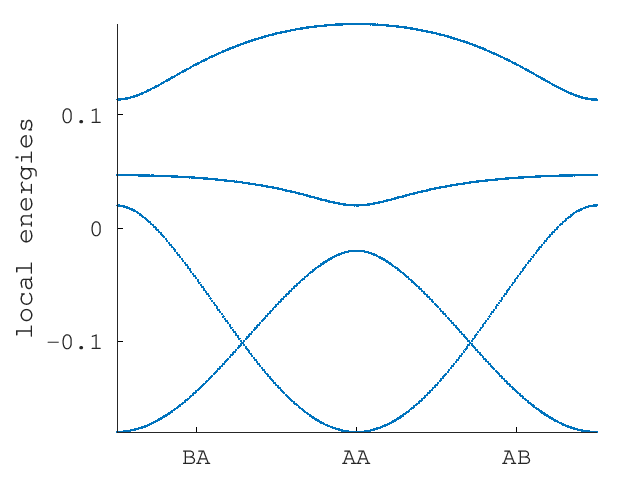}(a)
		\includegraphics[width=0.3\columnwidth]{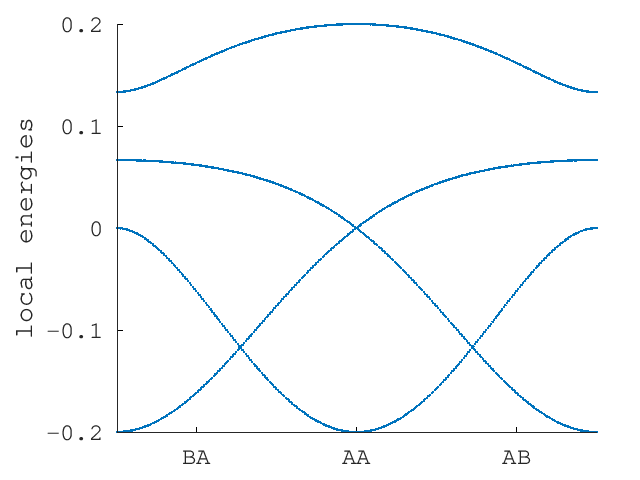}(b)
		\includegraphics[width=0.3\columnwidth]{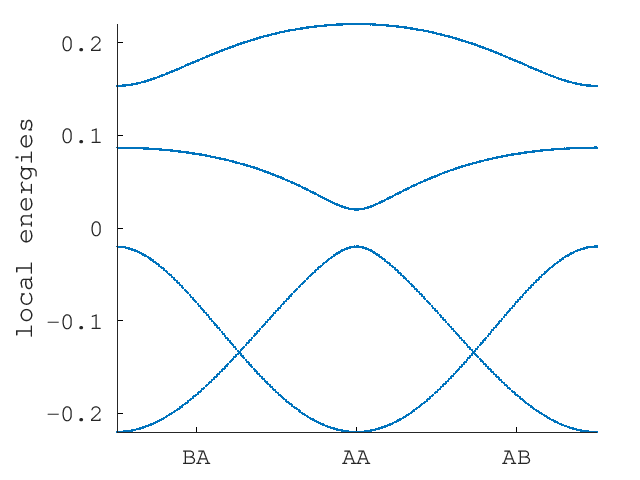}(c)
		\includegraphics[width=0.3\columnwidth]{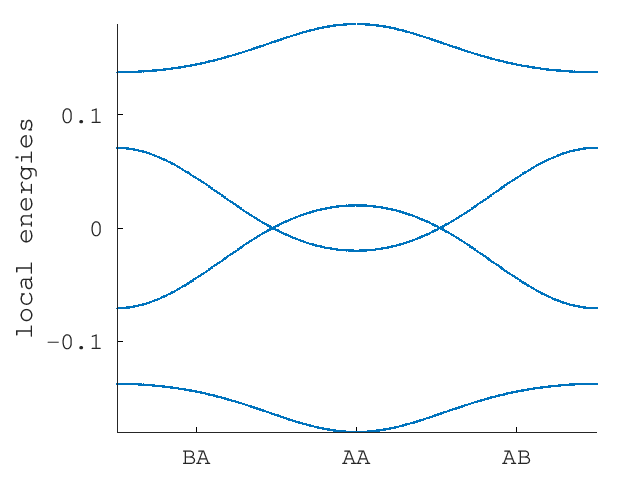}(d)
		\includegraphics[width=0.3\columnwidth]{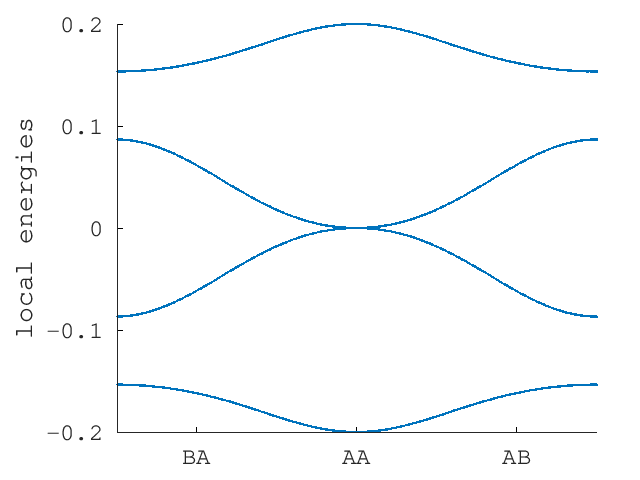}(e)
		\includegraphics[width=0.3\columnwidth]{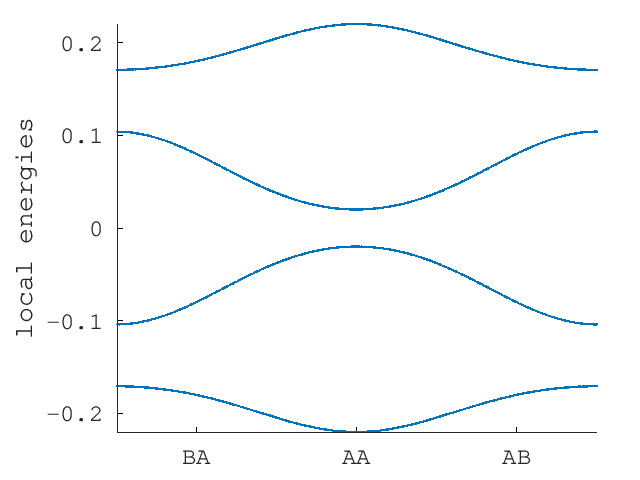}(f)
	\end{center}
	\caption{Spectra as a function of moir\'e position used in computing topological invariants.  (a) ZZ spectrum at $q_y < q_c$, yielding index 1.  (b) ZZ spectrum at $q_y = q_c$ where the gap closes on $AA$. (c) ZZ spectrum at $q_y > q_c$ with index 0.  (d) AC spectrum at $q_x = .1q_c$ and $q_y < q_c$, yielding index 1.  (e) AC spectrum at $\sqrt{q_x^2 + q_y^2} = q_c^2$ where the gap closes on $AA$.  (f) AC spectrum at $q_y > q_c$, index 0.}
	\label{invariant}
\end{figure}

A topological invariant classifying spectra as a function of $k_y$ is calculated over the local spatially varying spectra.  Figure~\ref{AC_local} shows that such an invariant cannot be calculated for the local AC Hamiltonian at $k_x=0$ because the spectrum is never gapped within the critical ring.  ZZ contains gapped regions within the critical ring where an invariant can be calculated, and likewise AC with a small nonzero $q_x$ to lift the gap closures allows an invariant to be calculated.  The spectra for which we calculate an invariant are shown in Figure~\ref{invariant}.  The $\mathbb{Z}_2$ invariant itself is a Wilson loop
$$ \frac{1}{\pi}\prod_n U(x_n)^\dagger U(x_{n+1}) $$
where $U(x)$ has as columns the two occupied states below the gap of $\mathcal{H}(x) = \gamma_{15} q_{15} + \gamma_{25} q_{25} + \boldsymbol{\nu}(x)$ and $x_n$ ranges from $0$ to $L$.

\section{Jackiw-Rebbi analysis}

The gap closures in local spectra correspond to mass inversions that localize zero modes found in the full moir\'e spectrum.  The critical ring localizes a zero mode on $AA$ stacking as $\nu_{14}$ changes sign, and the $AB/BA$-peaked modes localize where the combined mass $\sqrt{\nu_4^2 + \nu_{14}^2}$ crosses $\sqrt{\nu_{32}^2 + (v_F q_{25})^2}$.  To write an effective solution for the $AA$-peaked modes, we begin with the Dirac equation at zero energy
\begin{equation*}
	i\tilde{\bf t}_x \frac{\partial}{\partial x}\psi = -(q_y \tilde{\bf t}_y + \boldsymbol{\nu}(x))\psi\\
\end{equation*}
We can transform this equation to center on eigenstates of arbitrary $q_x$ by taking $\psi \rightarrow e^{-iq_x}\psi$.
\begin{align*}
	i\tilde{\bf t}_x \frac{\partial}{\partial x}e^{-iq_x}\psi = & -(q_y \tilde{\bf t}_y + \boldsymbol{\nu}(x))e^{-iq_x}\psi\\
	i\tilde{\bf t}_x \frac{\partial}{\partial x}\psi = & -(q_x \tilde{\bf t}_x + q_y \tilde{\bf t}_y + \boldsymbol{\nu}(x))\psi\\
	& \equiv -M(x)\psi
\end{align*}
Here $M(x)$ is the spatially varying local Hamiltonian at a fixed $q_x$, $q_y$.  Now we multiply both sides by the inverse of $i\tilde{\bf t}_x$ to obtain the differential equation
\begin{align}
	\label{diff_eq}
	\frac{\partial \psi}{\partial x} & = \frac{i}{v_F^2}\tilde{\bf t}_x M(x)\psi
\end{align}

\begin{figure}[!h]
	\begin{center}
		\includegraphics[width=0.3\columnwidth]{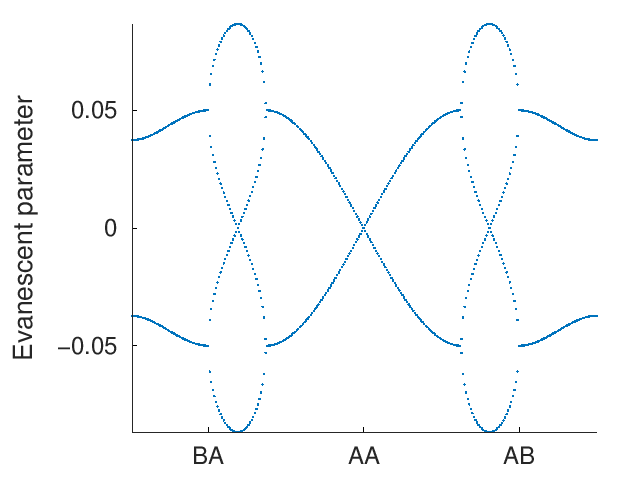}(a) 
		\includegraphics[width=0.3\columnwidth]{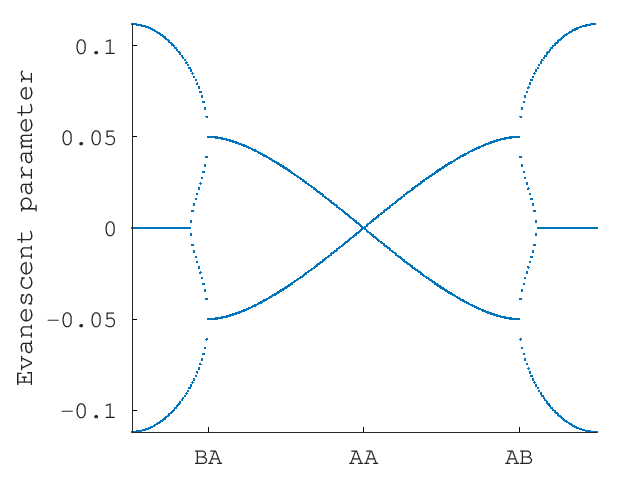}(b) 
	\end{center}
	\caption{Real part of the RHS eigenvalues of Equation~\ref{diff_eq} for (a) AC and (b) ZZ, describing evanescent growth and decay in the solutions.}
	\label{real_spectra}
\end{figure}

Assuming the moir\'e period is long, an operator varying at the rate of $\boldsymbol{\nu}(x)$ approximately commutes with the differential operator, with a neglibile commutator of order $1/L$.  Therefore we can diagonalize $M(x)$ without significantly altering the left hand side.  The real part of the eigenvalues of the right hand side describe evanescent growth and decay in decoupled solutions, so we look for branches crossing zero from positive to negative to describe localized peaks.  A zero eigenvalue of $M(x)$ automatically guarantees one of $\tilde{\bf t}_x M(x)$ because $M(x)$ annihilates the corresponding spinor, and these account for all the zero eigenvalues of the RHS because $\tilde{\bf t}_x$ is invertible.  Furthermore, the real part of the RHS eigenvalues are independent of $q_x$
\begin{align*}
	\frac{\partial \psi}{\partial x} & = \frac{i}{v_F^2}\tilde{\bf t}_x(q_x \tilde{\bf t}_x + q_y \tilde{\bf t}_y + \boldsymbol{\nu}(x))\psi \\
	& = \frac{i}{v_F^2}(q_x + \tilde{\bf t}_x(q_y \tilde{\bf t}_y + \boldsymbol{\nu}(x)))\psi 
\end{align*}
because commuting term $\frac{i}{v_F^2}q_x$ adds a pure imaginary term to the RHS eigenvalues.  We can conclude that at fixed $k_y$, a zero eigenvalue of $M(x)$ at {\it any} $q_x$ guarantees a critical point in the evanescent part of the solution.  In the gapped region of AC, this implies critical points in $AA$ as well as the drifting positions $\sqrt{\nu_4^2 + \nu_{14}^2} - \sqrt{\nu_{32}^2 + (v_F q_{25})^2} = 0$.  For ZZ, there must be critical points in $AA$, accompanied by a range at zero near $SP$ due the $SP$ cones crossing the gap.  Plots of the spectra (Figure~\ref{real_spectra}) confirm this and show that there are always branches changing from positive to negative as desired for a peaked solution.

\section{Carrier densities}

The physical measure of electron states in the weakly dispersing manifolds can be determined by the width of the critical ring $2q_c = 2t'/(\hbar v_F)$ with $\hbar v_F = \sqrt{3}at/2$ and the area of the supercell $A = \frac{\sqrt{3}}{2}aL$.  The spacing of states across the critical ring in $k$ space is $2\pi/L_y$, so that the number of states across the critical ring is $4 L_y t'/(2\pi\hbar v_F)$ which includes a factor of two for spin.  Dividing by area and letting $t'/t = 0.1$, $a = 1.42\times 10^{-8}$ cm, and $L = 1800a$, we obtain a density of
$$ \frac{4 t'}{2\pi\sqrt{3}t/2}\frac{1}{(\sqrt{3}aL/2)} = 2.33 \times 10^{11} e/\text{cm}^2$$
per band.  ZZ has the measure of four ring-width bands split across two valleys, and AC has eight bands split between four $AA$-peaked and four $AB/BA$-peaked.  Summing over four bands we obtain a measure of 
$$ 9.32 \times 10^{11} e/\text{cm}^2 $$
for each type of mode ($AA$ or $AB/BA$).  This is about $1/7$ the density required to fill four bands in magic angle twisted bilayer graphene, $n = 6.4 \times 10^{12} \text{cm}^{-2}$ \cite{Jarillo-Herrero}.

In terms of the strain $u$, which is about $\sqrt{3}a/L$, the carrier density to fill four bands is
$$ 4\frac{2t'/t}{\sqrt{3}/2}\frac{2u}{(3\pi a^2)} = (9.7 \times 10^{14})u \; e/\text{cm}^2$$
For $L=1800$, the strain about $u=.001$.  If the strain is increased to the order of $u=.01$, the corresponding carrier density is $9.7 \times 10^{12} \; e/\text{cm}^2$.

\end{document}